\documentclass{article}

\usepackage[a4paper, margin=1in]{geometry}
\usepackage{authblk}
\usepackage{amsmath}
\usepackage[running]{lineno}
\usepackage{graphicx}
\usepackage{epsfig}
\usepackage{float}
\usepackage{enumitem}

\usepackage[dvipsnames,svgnames,x11names]{xcolor}

\begin{document}
\title{Layer thickness dependent band gap\\of MBE grown single- to few-layer MoS$_{2}$}
\author[1]{Maciej Bazarnik\thanks{mbazarni@uni-muenster.de}}
\author[2]{Thorsten Deilmann}
\author[1,3]{Marta Przychodnia}
\author[1]{Anika Schlenhoff}
\affil[1]{Institute of Physics, University of Münster, Münster, Germany}
\affil[2]{Institute of Solid State Theory, University of Münster, Germany}
\affil[3]{Institute of Physics, Poznan University of Technology, Poland}
\maketitle
\begin{abstract}
In light of the rise of transition metal dichalcogenides as 2D semiconductors for device applications, band engineering becomes very important from an application point of view.
In many of these materials, such as the canonical example of MoS$_{2}$, the semiconductor band gap depends on the layer number. 
It changes from indirect to direct as it evolves from a bulk semiconductor to a monolayer.
Interestingly, it was predicted and experimentally confirmed that, by thinning the material from bulk to a bilayer, the indirect transition shows a strong blue-shift.
Here, we present the results of scanning tunnelling spectroscopy measurements on MoS$_{2}$ that has been grown \textit{in situ} via molecular beam epitaxy on graphene on Ir(111) at thicknesses ranging from 1 to 5 layers.
We find a drastic decrease of the band gap with increasing layer number, to values even below the band gap in bulk.
We also observe that the pinning of the conduction band vanishes above 4 layers.
Comparing our experimental data with density functional theory and \textit{GW} calculations indicates that an additional screening is introduced by the sample growth conditions.
\end{abstract}

%
\section*{Introduction}
Since the discovery of two-dimensional (2D) transition metal dichalcogenide (TMD) semiconductors, intensive work has been devoted to the control of their electronic and optical properties~\cite{wang_electronics_2012, chhowalla_chemistry_2013, butler_progress_2013}.
Being atomically thin, they have potential applications for nanoelectronic devices as well as transparent flexible electronics, similar to graphene.
Unlike graphene being a semimetal, they show a broad range of semiconducting band gaps achievable by selecting appropriate 2D binary compounds.
Moreover, 2D TMDs exhibit exotic properties such as indirect-to-direct band gap transition with decreasing number of atomic layers~\cite{splendiani_emerging_2010}, field-induced transport with high on–off ratios~\cite{radisavljevic_single-layer_2011, yin_single-layer_2012}, strong photovoltaic responses~\cite{bernardi_extraordinary_2013, britnell_strong_2013}, and interesting valleytronics phenomena~\cite{xiao_coupled_2012}.
Among the most interesting properties of TMD materials is the tunability of their electronic structures.
For instance, additional electronic states located in the band gap of the host material can be induced by the presence of defects such as dopants, vacancies, lattice antisites, etc.~\cite{fuhr_scanning_2004, zou_predicting_2013, zhou_intrinsic_2013, yue_functionalization_2013}.
The band gap itself is tunable  with layer thickness~\cite{jin_direct_2013, zhang_direct_2014-1, komsa_effects_2012, kuc_influence_2011, huang_bandgap_2015, bradley_probing_2015, trainer_inter-layer_2017, murray_comprehensive_2019},  stress~\cite{castellanos-gomez_local_2013, conley_bandgap_2013, feng_strain-engineered_2012}, by edge-dependent and grain boundary induced semiconducting-to-metallic transitions~\cite{pan_edge-dependent_2012, liu_point_2016, yan_charging_2018, wang_bandgap_2018, van_efferen_modulated_2024}, doping~\cite{efferen_metal-insulator_2022}, as well as by transformations between different stacking geometries~\cite{mattheiss_band_1973,lin_atomic_2014}.
While a small band gap in graphene (a few hundred meV) can be opened by stress and other methods~\cite{balog_bandgap_2010, xue_scanning_2011, warmuth_bandgap_graphene_2016}, the band gaps in 2D TMD materials vary significantly from zero to a few eV.
Correspondingly, the properties change from semiconducting to metallic and even superconducting, which allows a wide range of applications in a multitude of nanoscale devices.
To exploit all these extraordinary properties ultra clean thin TMDs are required, such as those achievable by molecular beam epitaxy (MBE)~\cite{hall_molecular_2018}.
Moreover, the growth should be carried out on the target surface that is planned either for further experiments or application since transfer methods can induce more defects or leave residue~\cite{fournel_molecular_2025}.
However, like other 2D materials, TMDs show a very strong variation of properties when placed on different substrates~\cite{bradley_probing_2015, murray_comprehensive_2019, efferen_metal-insulator_2022}.
Graphene on Ir(111) (Gr@Ir(111)) has been proven as a very inert substrate for the growth of molybdenum disulfide (MoS$_2$), enabling the experimental observation of excitons with comparably long lifetimes, indicating weak electronic coupling to graphene~\cite{ehlen_narrow_2018}.
Moreover, it has been suggested that high quality multilayers can be achieved by a series of repetitive depositions at room temperature and subsequent annealings~\cite{hall_molecular_2018}.
So far, most work on MoS$_2$ has focused on an 1~-~3\,Layer (L) thickness regime~\cite{bradley_probing_2015, trainer_inter-layer_2017, murray_comprehensive_2019, hall_molecular_2018}.
Here, we expand this regime to thicknesses ranging from a monolayer up to five layers, realized via \textit{in situ} MBE growth of MoS$_2$ on Gr@Ir(111). 
By employing scanning tunneling microscopy and spectroscopy (STM and STS), we investigate the different layers in detail to reveal the layer-thickness dependent variations in electronic properties.

\section*{Results and Discussion}
\begin{figure}[t]
\begin{center}
\includegraphics[width=\columnwidth]{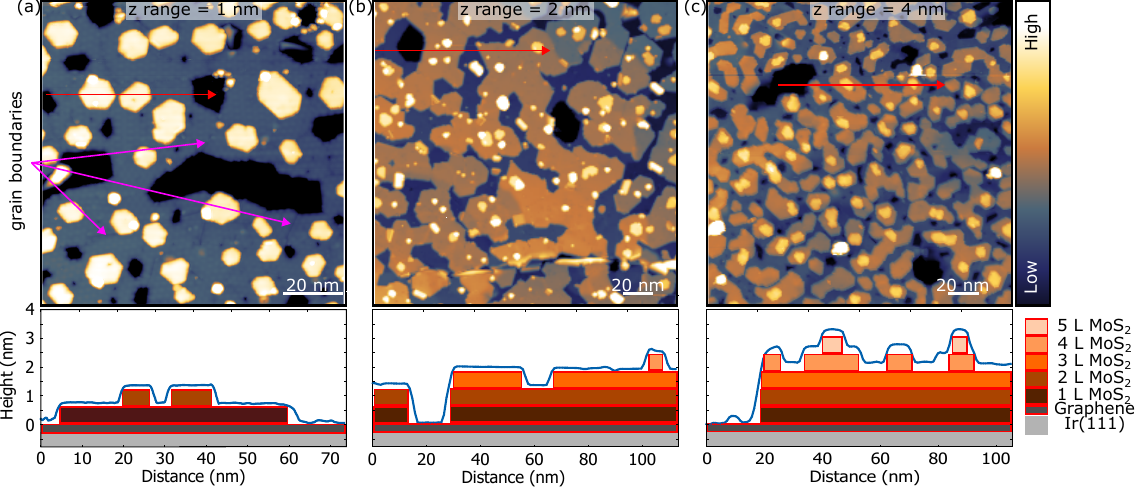}
\caption{\textbf{Surface morphology}
\textbf{(a)}~STM topography of a sample covered with monolayer and bilayer islands.
Multiple grain boundaries are visible in the monolayer.
\textbf{(b)}~STM topography of a sample covered with an almost fully closed bilayer and extensive trilayer on top.
Buried Ir steps are visible as small changes in the apparent height of the top layer.
\textbf{(c)}~STM topography of a sample with an almost closed trilayer and additional four and five layer islands on top.
Height profiles taken along the arrows marked in red are shown below each panel with a color coded sketch of the investigated system. 
Tunneling parameters: $I_t$~=~1\,nA, $U$~=~2\,V.
}
\label{fig_structure}
\end{center}
\end{figure}

Our samples were MBE grown on an \textit{in-situ} prepared Gr@Ir(111) by deposition of Mo onto Gr@Ir(111) at room temperature in a partial sulfur pressure.
Further, the sulfur atmosphere was maintained during the post deposition annealing at 1125\,K.
Each of our samples was grown in a single attempt.
This optimised the time and reduced the desorption of sulfur from edges. 
Figure~\ref{fig_structure} shows typical STM topographies of samples covered with (a) 1~-~2\,L, (b) 2~-~4\,L and (c) 3~-~5\,L MoS$_2$.
The height profiles below each panel are overlaid with a sketch of the multilayer composition.
In the lowest coverage range investigated, the monolayer covers nearly 80~\% of graphene, while part of it is already covered with second layer islands.
There are multiple domain boundaries visible in the monolayer indicating numerous nucleation sites and resulting in rotational domains of the monolayer.
Bilayer growth nucleates predominantly at defects and grain boundaries of the monolayer. In the latter case, the second layer adapts the configuration of one side of the grain boundary, leading to a rotation between the first and the second layer on the opposite side of the monolayer’s grain boundary.
However, no systematic spectroscopic difference is observed between second layer islands of different orientation.
In medium coverage regime (Figure~\ref{fig_structure}(b)), the bilayer covers almost all of the monolayer.
The trilayer forms an extended film, with a rather irregular edge structure indicating growth from multiple nucleation sites.
The four layer islands are relatively small with some excess amorphous clusters assigned as excess Mo~\cite{hall_molecular_2018}.
The irregular steps of Ir(111) below the film are visible as small changes to the apparent height of the top layer.
These steps form during the post deposition annealing process.
However, the crystal uniformity is easily restored by a few cleaning cycles involving sputtering and annealing.
In the highest coverage regime (Figure~\ref{fig_structure}(c)), the trilayer is almost completely closed and four and five layer islands are formed on top.
Since there is no strain in the initial monolayer, the epitaxial growth should continue without relaxation.
A continued layer-by-layer growth is expected~\cite{Liu_epitaxy_2020}.
Instead, the bilayer already starts growing preferably before the monolayer is fully closed, indicating that the typical growth kinetics are hindered from the various available nucleation sites.
This trend extends to 3 - 5~L and in effect we observe a Stranski–Krastanov type of growth~\cite{hall_molecular_2018, Wouter2020}.
Evidently, some areas of graphene remain uncovered even when an almost closed three layer film is covering the sample.
In samples with more than two layers, these areas are limited to less than 5\% of the surface coverage.

\begin{figure}[ht!]
\begin{center}
\includegraphics[width=0.5\columnwidth]{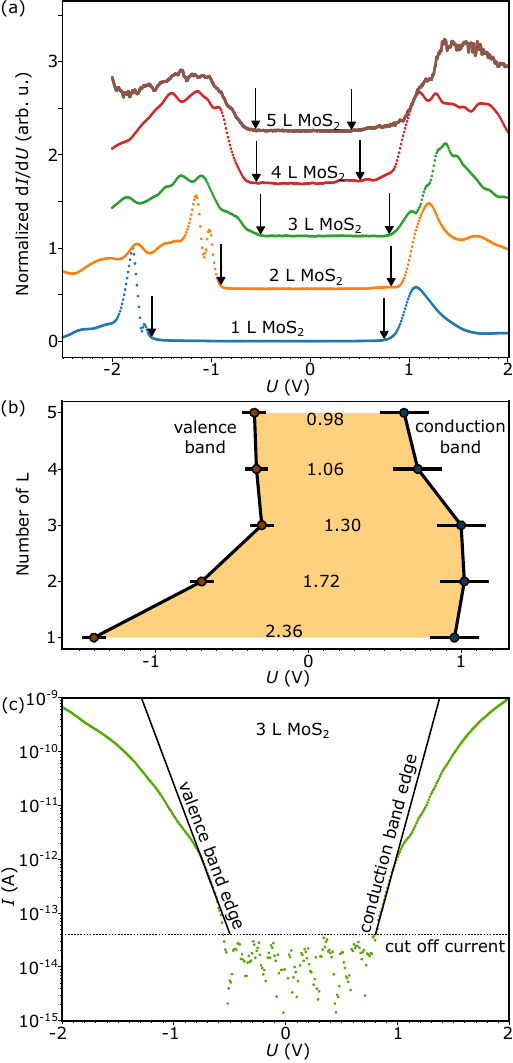}
\caption{\textbf{Electronic structure}
\textbf{(a)}~Normalized (to 1) d$I$/d$U(U)$ data averaged over several curves for 1~-~5\,L of MoS$_2$.
The valence band and conduction band edge are indicated with arrows.
Curves are vertically offset by 0.6 for clarity.
\textbf{(b)}~Extracted band edge positions and determined band gaps. Points connection and coloured area marked to increase the results' visibility.
\textbf{(c)}~An exemplary semi-logarithmic $I(U)$ curve for 3\,L MoS$_2$.
Cut off current (defined as noise level of the preamplifier) and fitted valence and conductance band edges.
The exact values marked in \textbf{(a)} and \textbf{(b)} are the crossing points of the fit and the cut off current.
Stabilization parameters: $I_t$~=~1\,nA, $U$~=~2\,V.
}
\label{fig_sts}
\end{center}
\end{figure}

Now, we turn to the electronic characterization of these layers.
To determine the valence band and conduction band edge we performed STS measurements. 
Recently, a comprehensive STS study of monolayer and bilayer MoS$_2$ grown on Gr@Ir(111) has been performed using the constant current spectroscopy mode~\cite{murray_comprehensive_2019}, yielding detailed results.
Unfortunately, this constant current mode cannot be conducted on thicker semiconducting samples.
With increasing layer thickness and corresponding increasing tip-graphene separation, the electron tunneling to the graphene within the MoS$_2$ band gap is progressively suppressed.
As a consequence, when regulating on a constant current in the spectroscopy mode, with an applied bias corresponding to an in-gap energy, the tip would further and further approach to the sample until it crashes into the MoS$_2$.
Therefore, for data consistency within the whole thickness regime, we deduced all the band gaps from constant height measurements.
Note, that all spectroscopy curves have been recorded far away from island edges and any visible defects to avoid any perturbations on the electronic properties.
Figure~\ref{fig_sts} summarizes our findings on the electronic properties and the band gap analysis.
Panel~(a) presents the differential conductance (d$I/$d$U(U)$) curves registered with a lock-in technique and normalized to 1. 
The extracted bang gaps are presented in panel~(b).
The band edges are deduced from the intersection of a linear fit to a logarithm of the tunneling current (as a function of bias) in proximity of the band gap and the cut off current (set to the noise level of the current preamplifier at 40 fA), as shown exemplary for the 3\,L MoS$_2$ in Fig.~\ref{fig_sts}(c).
The resulting valence and conduction band edges are indicated with arrows in the plots for all layers numbers.
The band edges of 1~-~3\,L are easily observable by an increase in the differential conductance in Fig.~\ref{fig_sts}(a), while for 4~-~5\,L the conduction band does not show a sharp increase in signal.
This observation in STS usually correlates with the band extrema being located apart from the $\Gamma$ point. 
On the other hand, the rise in the d$I/$d$U$ signal stemming from the onset of the valence band is clearly observable for all investigated systems.
As the number of layers increases, more maxima appear in the d$I/$d$U$ signal indicating a more complex structure for both occupied and unoccupied bands.
The band edge positions and the corresponding band gaps are shown in Fig.~\ref{fig_sts}(b).
The conduction band is found to be pinned for 1~-~3\,L while the valence band is shifted towards the Fermi level.
This finding contrasts with observations for chemical‑vapor‑deposition‑grown MoS$_2$ on graphite, where a pinning of the valence band and a downward energy shift of the conduction band were reported~\cite{huang_bandgap_2015}.
Contrary, for the 3~-~5\,L the valence band is pinned and the conduction band is shifting towards the Fermi level.
Overall, a significant red shift of the band gap from $(2.36 \pm 0.16)$\,eV to $(0.98 \pm 0.16)$\,eV is observed.
Note that the value of 0.98\,eV for the 5 L band gap is smaller than the value of 1.29\,eV expected for bulk MoS$_2$~\cite{textbook}.

Interestingly, 3 L exhibit a band gap that is within the margin of error equal to the bulk value.
Only by extending our thickness-dependent band gap analysis beyond the previously studied range of up to 3\,L, we experimentally reveal that the latter layer number is not sufficient for the development of a bulk-like band structure.

\begin{figure}[hbt!]
\begin{center}
\includegraphics[width=0.5\columnwidth]{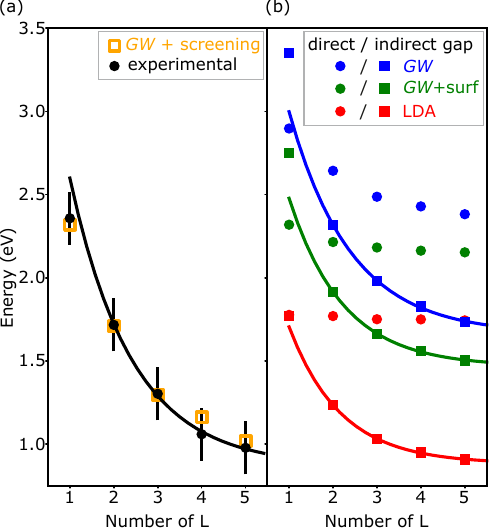}
\caption{\textbf{Comparison of thickness-dependent band gap evolution obtained from experiment and different theoretical models.}
\textbf{(a)}~The band gaps extracted from the experiments plotted together with band gaps obtained by adding the additional screening in \textit{GW} calculation.
An exponential decay fit to the experimental data for thicknesses 2~-~5\,L is overlaid.
\textbf{(b)}~Direct (circles) and indirect (squares) band gap as a function of film thickness calculated with different methods.
Exponential decay fits to band gap for thicknesses 2~-~5\,L obtained for all methods are included.
}
\label{fig_calculations}
\end{center}
\end{figure}

Figure~\ref{fig_calculations}(a) shows the experimentally obtained band gaps (black color) with an exponential decay fit  for 2~-~5\,L.
The fit yields a band gap of $(0.87\pm0.05)$\,eV for infinite layer numbers, which is about 0.1\,eV lower than the one observed at 5\,L.
The question arises whether the observed continuous redshift is an intrinsic property of the MoS$_2$. 
To answer this question we performed calculations within the local density approximation (LDA) of the density functional theory (DFT) and many-body perturbation theory in the \textit{GW} approximation for varying film thicknesses.
In a first step, we performed both LDA and \textit{GW} for the system in vacuum.
To simulate the dielectric properties of the substrate,
we have added corresponding dielectric model functions below the slabs~\cite{AuPTCDAAu}
(denoted as \textit{GW}+surf).
Figure~\ref{fig_calculations}(b) summarizes these model calculation.
The LDA result (red color) shows almost no change of the direct band gap as a function of MoS$_2$ layer thickness.
This is a consequence of the negligible hybridization at the $\pm$K point where the direct gap is located.
However, a decreasing tendency of the indirect band gap is found due to the hybridization at $\Lambda$ and $\Gamma$.
For the monolayer, the indirect and direct gaps are almost identical in energy, with the indirect one being smaller by about 5\,meV.
However, LDA is known to clearly underestimate band gaps in general and so it does for the monolayer MoS$_2$.
The discrepancy between the calculated band gap (1.77 eV) and the experimental result (2.36 eV) for the monolayer is the most obvious.
Even though this difference reduces up to five layers, with a calculated gap of 0.91\,eV compared to the experimentally determined value of 0.98\,eV, 
LDA does not reproduce the exponential decay well.

Typically, DFT band gap calculations are improved when considering a better theoretical footing.
One of today's most reliable results can be achieved by the \textit{GW} approximation.
For the monolayer we find a direct band gap of 2.89\,eV (blue color), and an indirect gap of 3.35\,eV.
In this case, the direct gap is in fact much smaller then the indirect one, and both type of gaps reduce with the number of layers at a rate similar to the experimentally observed one.
Despite this layer-thickness dependent trend being similar to the experimentally observed one, the size of the band gap is strongly overestimated.
The reason is the assumption, that the slab resides in vacuum.
In experiment, however, it is placed on a substrate
and thus we include the surface screening in our calculation (green color).
This approximation seems valid in the first place,
as no chemical hybridization of MoS$_2$ with the substrate is expected.
Here, the values for the band gap in the monolayer are very close to the experimental ones,
but the trend with increasing layer number is less steep.
The indirect band gap, which is the smaller one for two and more layers,
drops distinctly less compared to the experimental data.
For better comparison, the calculated data are fitted by an exponential decay function, including the indirect band gaps for 2~-~5\,L.
The fittings demonstrate that all our calculations show a continuous drop that would even continue beyond the five layer results, however the band gap value levels out faster than the experimentally observed trend shows.

Besides the hybridization, another crucial property is neglected in our calculations so far: a potential doping which introduces additional metallicity in the system.
Since we do not have direct experimental access to the amount of changing doping, we have to approximate this screening.
Such \textit{GW} calculations with additionally screening are shown in orange in Figure~\ref{fig_calculations}(a) and are fitted to match the experimental data.
We find, that the introduced screening has to be increased with the number of layers, starting with 1\% for two layers, 10\% for three, 20\% for four up to almost 100\% for five layers.

The physical origin of this additional screening cannot be deduced from our simulations.
Even though from the theoretical point of view this additional screening could be caused by intercalation of Mo or S in between layers during growth, our experimental data invalidates this hypothesis.
Additional material in between layers would lead to an increase in the apparent height.
However, in experiment we observe that the 4 L and 5 L are actually marginally lower in the height profile.
While we position our tip above ``clean'' regions,
we are only able to see the respective topmost layer.
Since we have numerous structural defects in sub-surface layers caused by multiple nucleation sites as described above, we expect buried grain boundaries and stacking faults in the multilayer system.
Stacking faults such as found in twisted bilayer of MoS$_2$ are known to modify the band gap~\cite{Zhang_twisted_bilayer_2020}.
Moreover, it is known that grain boundaries in monolayer MoS$_2$ on Gr@Ir(111) show metallic states~\cite{van_efferen_modulated_2024}, while grain boundaries in the topmost layer of an 1~-~3\,L system on graphite have been found to reduce the band gap by up to 0.85~eV~\cite{huang_bandgap_2015}.
Consequently, it is reasonable to assume that these buried defects are the reason for the observed reduced band gap in our multilayer system compared to the pristine bulk value.
Furthermore, with increasing layer number the total amount of buried defects, potentially residing in each layer, increases.
Hence, an increasing additional screening for thicker slabs, as required in the model calculations to reproduce the experimental trend, is plausible.

\section*{Conclusions}
We have performed a combined STM/STS and DFT/\textit{GW} study of the influence of layer thickness on the MoS$_2$ band gap in the 1 - 5~L regime.
The feasibility of single-cycle MBE growth of several layers MoS$_2$ on a Gr@Ir(111) substrate has been demonstrated. 
Samples prepared in this way exhibit a Stranski–Krastanov type of growth and a high density of grain boundaries and rotational domains.
Both the calculations and experimental observations indicate that a thickness of 5~L is still insufficient to develop a `bulk-like' band gap comparable to that of an infinitely thick film. 
Interestingly, in our experiments the 3~L MoS$_2$ coincidentally exhibit a band gap equal to the value expected for bulk MoS$_2$.
By extending this previously studied thickness range, we experimentally observe a pronounced red‑shift of the band gap, decreasing from $(2.36 \pm 0.16)$\,eV for 1~L to $(0.98 \pm 0.16)$\,eV for 5~L, reaching values that are even lower than the known bulk MoS$_2$ gap.
This thickness-dependent evolution of the band gap is reproduced by \textit{GW} calculations including an additional screening that scales with the film thickness.
We suggest that the driving factor behind this extra screening is the growth mode, which introduces surface and buried irregularities such as grain boundaries, rotational domains, and stacking faults.

\section*{Methods}
Ir(111) was cleaned by repeated cycles of Ar$^{+}$ sputtering ($1$\,kV, $5$E$-6$\,mbar), annealing at temperatures ranging from $T=900$\,K to $1500$\,K in the presence of oxygen and a flash annealing at $T\approx 1500$\,K.
The Gr layer was grown \textit{in situ} on Ir(111) by thermal decomposition of ethylene molecules following the procedure described in ref.~\cite{NDiaye2008}. 
MoS$_2$ was grown by MBE deposition of Mo out of an e-beam evaporator onto a substrate held at room temperature and in partial pressure of S ($1$E$-7$\,mbar), followed by a subsequent annealing to 1170\,K while maintaining the S atmosphere.
S was supplied by thermal decomposition of FeS$_2$ in a Knudsen cell held at between 500\,K and 600\,K, while the partial pressure in the chamber was monitored by a mass spectrometer.
During growth, the background pressure was better than $1$E$-9$\,mbar.
After preparation, samples were transferred \textit{in vacuo} to the STM setup and cooled down to the measurement temperature of $T=5.5$\,K.
A lock-in detection technique was used to obtain d$I$/d$U$ point spectroscopy data adding a small AC modulation $U_{\rm {Pk-Pk}}=20$\,mV to the bias voltage~$U$ at $f=3846$\,Hz. 
Electrochemically etched tungsten tips, cleaned by standard \textit{in vacuo} flash annealing at $T\approx 2500$\,K, were used.
All data has been processed using self-written python code and Gwyddion~\cite{Necas2012} software.

Our theoretical approach for the calculation of MoS$_2$ has been discussed in detail for the monolayer in Ref.~\cite{GdWTMDC}.
For the current work, we employ one to five layers MoS$_2$ in the $2H$ phase.
We employ the defect-free experimental structure
and use the LDA (local density approximation) of the density functional theory, as well as the $GdW$ approximation of many-body perturbation theory.
Herein the screening is simulated by atom-centered model functions~\cite{GdW}, which are parameterized by ab-initio calculations with the random phase approximation.
For the additional screening of the surface we follow our approach~\cite{AuPTCDAAu} in which we add additional metallic model functions at the position of the substrate underneath the corresponding slab.
For the assumed additional doping we add similar, but weighted, metallic model functions at the atom positions of the corresponding slabs.
The weights of 1\% for two layers, 10\% for three, 20\% for four and almost 100\% for five layers have been roughly guessed by the comparison to the experimental findings.

\section*{Acknowledgments}
T.D. acknowledges financial support from the Deutsche Forschungsgemeinschaft (DFG, German Research Foundation) through Project No. 426726249 (DE 2749/2-1 and DE 2749/2-2).
The authors gratefully acknowledge the Gauss Centre for Supercomputing e.V. (www.gauss-centre.eu) for funding this project by providing computing time through the John von Neumann Institute for Computing (NIC) on the GCS Supercomputer JUWELS~\cite{JUWELS} at Jülich Supercomputing Centre (JSC).

\section*{Author Contributions}
AS conceived the experiments and lead the project.
MB and MP performed the experiments and analyzed the data.
TD performed the calculations and analyzed the data.
All authors discussed the data and wrote the manuscript.

\bibliographystyle{naturemag}
\bibliography{LayerdSTSMoS2}

\end{document}